\documentclass{article}
\usepackage{amsmath}
\usepackage[dvips]{graphics}
\begin{document}

\title{{\bf Exactly Solvable Model of Inergodic Spin Glass}}
\author{P. N. Timonin \\ 
Physics Research Institute at Rostov State University,\\344090,
Rostov-on-Don, Russia\\e- mail: timonin@icomm.ru}
\maketitle

\begin{abstract}
A mean-field model of Ising spin glass with the Hamiltonian being a sum of
the infinite-range ferromagnetic and random antiferromagnetic interactions
has been studied. It is shown that this model has phase transition in
external magnetic field into inergodic spin glass phase with a number of
metastable states. The thermodynamic properties of metastable states are
studied at $T=0$ and near the transition. The relations between the
characteristics of slow nonequilibrium processes in spin glass phase (such
as hysteresis loop form, thermo-remanent and isothermal remanent
magnetizations, field-cooled and zero-field-cooled thermodynamic quantities)
and thermodynamic parameters of metastable states are established.
\end{abstract}

\section{Introduction}

The possibility of existence of spin glass phases as specific thermodynamic
states in solid solutions of ferromagnets and antiferromagnets has been
first suggested in pioneering work by Edwards and Anderson \cite{1}.
According to\cite{1} , these phases are characterized by the appearance of
local spontaneous magnetic moments with chaotic orientations, which are
determined by a random distribution ferromagnetic and antiferromagnetic
interactions throughout the crystal. The first attempt to give the
quantitative description of spin glass transition was made by Sherrington
and Kirkpatrick (SK)\cite{2}, who considered a mean-field model with
infinite range random interaction. But the solution they got appeared to be
unstable in the glass phase for $T<T_{sg}$, $H<H_{AT}\sim \left(
T_{sg}-T\right) ^{3/2}$\cite{3}. The further attempts to describe the
thermodynamics of spin glass phase below the Almeida-Thouless (AT) line, $%
H<H_{AT}$, resulted in the construction of 'replica symmetry breaking'
scheme by Parisi \cite{4} being the procedure of the analytical continuation
in the replica method used in the studies of the SK model. Now it is common
belief that Parisi solution is stable below AT-line and it is the basic
result in the spin glass theory. With some reservations concerning the
mathematical foundations of the replica method and the suggested procedure
of the analytical continuation, one may consider the Parisi's solution as
the first exact description of the thermodynamics of inergodic spin glass
phase.

Later, the other exactly solvable spin glass models have been studied with
the use of the replica method (the vector models \cite[6]{5}, $p$-spin
spherical models \cite[8]{7}) and without it (ordinary spherical spin glass 
\cite{9}, Bethe lattice spin glass\cite{10}). Let us also notice the study
of the mean-field equations for the local magnetic moments of the SK model 
\cite{11}. In these studies the equilibrium thermodynamic parameters
averaged over random interactions were obtained in the framework of the
standard statistical mechanics. Still these results appear to be
insufficient for the description of properties of real spin glasses. The
reason lies in the known inergodicity of the spin glass phases \cite{12}, i.
e. the existence a large number of metastable states in these phases. So the
experiments give the physical quantities proper to one of these states, in
which crystal comes depending on a regime and a sequence of cooling and
application of magnetic field \cite[17]{13}. Meanwhile, the equilibrium
quantities refer only to the state with the lowest thermodynamic potential
and could be obtained after sufficiently large observation time such that
crystal could come to the lowest equilibrium state when field and/or
temperature are changed. As the barriers between metastable states are
macroscopic (divergent as $N\rightarrow \infty $), the corresponding
relaxation times are generally astronomically large. Thus crystal can stay
in the initial state while his potential becomes larger then some other
state.

The situation can be elucidated by considering the uniaxial (Ising)
ferromagnet being the simplest inergodic system below $T_c$. In the fields,
smaller then coercive one, it has two stable states: equilibrium state with
magnetization $m$ parallel to $H$, and metastable one having $m$
antiparallel $H$ and greater potential. The standard result of the
equilibrium statistical mechanics for the dependency $m$ on $H$ in this case
is a function with a jump at $H=0$ corresponding to the equilibrium states,
while in the real experiments the hysteresis loop is observed, in which
metastable states are also present. Thus the description of real experiments
needs also the description of the properties of metastable states. In this
example this is trivial at least in the mean-field approximation, but in
spin glasses the description of the properties of a large number of
metastable states is rather difficult task. In particular, the description
of magnetic properties of a spin glass phase must include a set of functions 
$m(H)$, corresponding to various metastable states, and their lines on the $%
m-H$ plane would fill some region around the origin. In a rather simplified
form, the theoretical problem is to determine the boundaries of this
inergodic region and the values of various thermodynamic quantities for all
points inside it. Such theoretical results could describe a number of slow
nonequilibrium processes in the inergodic spin glass phases.

Meanwhile, all data about metastable states obtained via Parisi's ansatz
come to the probability distribution of the 'overlaps' of magnetizations in
various metastable states \cite{13}. Now it is not clear how this
distribution can describe the real experiments. Also there was not got any
information on the properties of metastable states of the SK model from the
TAP equations \cite{11}, it was only established that their number is
exponentially large \cite{14}.

The studies of the other spin glass models have not also result in the
elucidation of the physical characteristics of metastable states. The only
exception is the Ising spin glass on the Bethe lattice for which the
numerical study of the internal field distribution has explicitly shown the
existence of a number of metastable states at $T=0$\cite{10}.

Generally speaking, the study of the properties of metastable states is not
necessary for a description of experiments in inergodic systems as it could
be obtained from the study of nonequilibrium dynamics which incorporates
automatically the effects of their existence. Thus one can get the
hysteresis loop as a reaction on the large slow varying field in the dynamic
treatment of the uniaxial ferromagnet. But in doing this one must eliminate
large unobservable times of the relaxation between metastable states. Such
elimination procedure has been developed in the study of Langevin dynamics
of SK model in zero field \cite{15}, which help to establish the difference
between the unobservable equilibrium susceptibility and those in the local
metastable states measured at real times smaller than that of relaxation
over macroscopic barriers. Still now it is not clear how this procedure
could be generalized to describe in the SK model the reaction on the
alternating field of finite amplitude. Now the inergodic effects in finite
fields were described only in the simplest spherical spin glass model having
two stable states \cite{16}.

To resume we may say that the spin glass models considered until now are not
sufficiently simple to obtain the thermodynamic properties of metastable
states needed for description of inevitably nonequilibrium processes in spin
glasses. Hence it seems worthwhile to find and to study more simple models
allowing for more complete description of physical properties in the
inergodic glass phases. Here we present one such model which allows the
analytical description of thermodynamic characteristics of all metastable
states at $T=0$ and near the transition point.

\section{Hamiltonian of the model and its properties at $T=0$.}

The most simple treatment of metastable states is possible in the framework
of mean-field models with infinite range interactions when it comes to the
determination of local minimums thermodynamic potential. Also the random
spin glass interaction should be sufficiently simple in all its realizations
to make possible an analytical treatment. In Ising spin glasses the glass
transition results from the competition between ferromagnetic and
antiferromagnetic interactions so the task is to imitate this competition
using simple infinite range interactions.

Let us consider the system having $N$ Ising spins $S_{i\alpha }=\pm 1$,
divided on $N_b$ blocks which consist of $N_s$ spins, $N=N_bN_s$. Here index 
$\alpha $ is the block number and index $i$ is the number of the spin inside
the block. So the magnetization of the block is 
\begin{equation}
\label{1}m_\alpha =N_s^{-1}\sum\limits_{i=1}^{N_s}S_{\alpha i} 
\end{equation}
and the total magnetization is 
\begin{equation}
\label{2}m_\alpha =N^{-1}\sum_{\alpha
=1}^{N_b}\sum\limits_{i=1}^{N_s}S_{\alpha i}=N_b^{-1}\sum_{\alpha
=1}^{N_b}m_\alpha 
\end{equation}
Let us also introduce the 'antiferromagnetic' order parameter 
\begin{equation}
\label{3}\mu _\alpha =m_\alpha -m 
\end{equation}
The model Hamiltonian is a sum of ferromagnetic and 'antiferromagnetic'
interactions and the external field term 
\begin{equation}
\label{4}{\cal H}=-\frac N2Jm^2-\frac{N_s}2J_1\sum_{\alpha =1}^{N_b}\mu
_\alpha ^2-NHm 
\end{equation}
Here $H$ is the homogeneous external field, $J>0$ - ferromagnetic exchange
integral with a fixed value, $J_1>0$- random 'antiferromagnetic' exchange
integral.

The term proportional to $J_1$ imitates the antiferromagnetic bonds
distributed throughout a crystal so we assume that $N_b>>1$. Also we assume
that $N_s$ diverges in thermodynamic limit ($N\rightarrow \infty $) in order
the mean-field approximation to be valid.

In general, for large $N_b$ the 'antiferromagnetic' term in Eq.\ref{4} gives
rise to a number of various types of ordering with $m=0$, so it would be
more correct to call it the 'glass' term. The more so that the block analog
of the Edwards-Anderson order parameter%
$$
q=N_b^{-1}\sum_{\alpha =1}^{N_b}m_\alpha ^2-m^2 
$$
can be represented as 
\begin{equation}
\label{5}q=N_b^{-1}\sum_{\alpha =1}^{N_b}\mu _\alpha ^2 
\end{equation}
Thus we may say that for $N_b>>1$ the model Hamiltonian, Eq.\ref{4},
describes the competition of ferromagnetic and 'glass' ordering.

For $N_s\rightarrow \infty $ it is easy to find the (nonequilibrium)
thermodynamic potential depending on $m_\alpha $:%
$$
F\left( {\bf m}\right) =H\left( {\bf m}\right) /N-TS\left( {\bf m}\right) 
$$
where $S\left( {\bf m}\right) $ is the entropy per spin:%
$$
S\left( {\bf m}\right) =N^{-1}\ln \left[ Tr\prod_{\alpha =1}^{N_b}\delta
_{N_sm_\alpha ,\sum\limits_iS_{i\alpha }}\right] \approx 
$$
\begin{equation}
\label{6}\approx N_b^{-1}\sum_{\alpha =1}^{N_b}\left[ \ln 2-\frac{1+m_\alpha 
}2\ln \left( 1+m_\alpha \right) -\frac{1-m_\alpha }2\ln \left( 1-m_\alpha
\right) \right] 
\end{equation}
For $N_s\rightarrow \infty $ the description of the equilibrium
thermodynamics comes to the finding of the lowest minimum of $F\left( {\bf m}%
\right) $ while the less deep minimums correspond to metastable states. The
equations defining the extrema of $F\left( {\bf m}\right) $ are: 
\begin{equation}
\label{7}Tarc\tanh m_\alpha +\left( J_1-J\right) m-J_1m_\alpha =H 
\end{equation}
The minimums correspond to the solutions of Eqs.\ref{7} with the positively
defined Hessian 
\begin{equation}
\label{8}\frac{\partial ^2F\left( {\bf m}\right) }{\partial m_\alpha
\partial m_\beta }=\left[ T/\left( 1-m_\alpha ^2\right) -J_1\right] \delta
_{\alpha \beta }+\left( J_1-J\right) /N 
\end{equation}
It is easy to show that this simple model is inergodic at $T=0$. In this
case Eqs.\ref{7} become 
\begin{equation}
\label{9}m_\alpha =sign\left[ H+\left( J-J_1\right) m+J_1m_\alpha \right] 
\end{equation}
When $\left| H\right| >\max \left( J,2J_1-J\right) $ Eqs.\ref{7} have unique
solution $m_\alpha =signH$, while at $\left| H\right| >\max \left(
J,2J_1-J\right) $ they also have a number of solutions with arbitrary signs
of $m_\alpha $ limited only by the condition 
\begin{equation}
\label{10}\left[ H+\left( J-J_1\right) m\right] ^2<J_1^2 
\end{equation}
All these solutions are stable thus corresponding to the metastable states
of the model. The total magnetization in these states acquires a set of
discrete values%
$$
m=\frac{2n}{N_b}-1 
$$
Here $n$ is the number of blocks with magnetization $m_\alpha =1$, $%
n=0,...,N_b$. There are $\binom{N_b}{n}$ states with a given $m$ which differs
by the permutations of $m_\alpha $. The total number of inhomogeneous
metastable states cold be up to $2^{N_b}-2$ for a given $H$.

The energy per spin in these states is determined by their magnetization%
$$
E=\left( J-J_1\right) m^2/2-mH-J_1/2 
$$
and the entropy, Eq.\ref{6} is zero.

The equilibrium magnetization corresponding to the states with minimal
energy is $m_{eq}\left( H\right) =signH$ for $J>J_1$ and 
\begin{equation}
\label{11}m_{eq}\left( H\right) =2\sum_{n=1}^{N_b-1}\frac n{N_b}\theta
\left( N_b^{-2}-\varepsilon _n^2\right) +sign\left( \varepsilon
_{N_b-1}-N_b^{-1}\right) 
\end{equation}
when $J<J_1$. Here $\varepsilon _n\equiv \frac H{J_1-J}-\frac{2n}{N_b}+1$,
and $\theta $ is the Haviside's step function. The field dependencies of
magnetization in the equilibrium and metastable states are shown in Fig.1. 
\begin{figure}
\centering
\includegraphics[116,619][468,733]{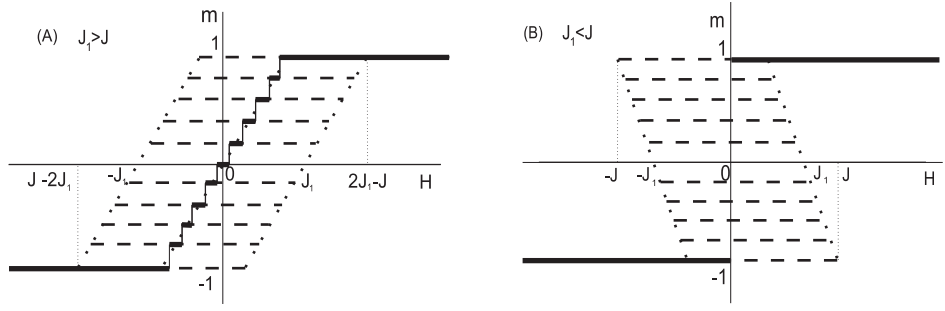}
\caption{Field dependencies of magnetization in
metastable states (dashed lines) and stable states (solid lines) at T=0. }
\end{figure}
The steps of the function $m_{eq}\left( H\right) $ for $J<J_1$demonstrate the
existence of the first order phase transitions at field values%
$$
H_n=\left( \frac{2n+1}{N_b}-1\right) \left( J_1-J\right) 
$$
$n=0,...,N_b$, at which the upturns of the block magnetizations take place.
Qualitatively, just this behavior of $m_{eq}\left( H\right) $ one may expect
in the glass phase of the magnet with equal concentrations of ferromagnetic
and antiferromagnetic bonds, while the case $J>J_1$ gives a picture proper
for a magnet with domination of ferromagnetic bonds. Further we will
consider the most interesting case when model Hamiltonian imitates a spin
glass, so we assume that probability distribution for $J_1$, $P(J_1)$ , is
zero for $J>J_1$ and has the form 
\begin{equation}
\label{12}P(J_1)=\theta \left( J_1-J\right) W\left( J_1-J\right) 
\end{equation}
In general, one may notice that averaging over random interaction is
superficial when all thermodynamic parameters can be obtained for every
random realization as experimental data are not usually averaged over a
number of samples. Still in more complex models it is often possible to find
only average equilibrium quantities. So it is interesting to compare them
with the corresponding quantities in the metastable states. Thus in present
model the magnetization in all states does not depend on the field and
magnetic susceptibility in them is zero at $T=0$. In the same time,
averaging of Eq.\ref{11} over $J_1$ gives for $N_b>>1$%
\begin{equation}
\label{13}\left\langle m_{eq}\left( H\right) \right\rangle
=H\int\limits_{\left| H\right| }^\infty \frac{dJ^{\prime }}{J^{\prime }}%
W\left( J^{\prime }\right) +signH\int\limits_0^{\left| H\right| }dJ^{\prime
}W\left( J^{\prime }\right) 
\end{equation}
\begin{equation}
\label{14}\left\langle \chi _{eq}\right\rangle =\frac{\partial \left\langle
m_{eq}\left( H\right) \right\rangle }{\partial H}=\int\limits_{\left|
H\right| }^\infty \frac{dJ^{\prime }}{J^{\prime }}W\left( J^{\prime }\right) 
\end{equation}
As one can see from Fig. 1(a), the non-zero $\left\langle \chi
_{eq}\right\rangle $ appears due to the series of transitions between
metastable states. Indeed the differentiation of Eq.\ref{11} gives 
\begin{equation}
\label{15}\chi _{eq}=\frac{\partial m_{eq}\left( H\right) }{\partial H}%
=\frac 2{N_b\left( J_1-J\right) }\sum_{n=1}^{N_b-1}\delta \left(
N_b^{-1}-\varepsilon _n\right) 
\end{equation}
When $N_b\rightarrow \infty $ Eq.\ref{15} becomes in the sense of
distributions 
\begin{equation}
\label{16}\chi _{eq}=\frac 1{\left( J_1-J\right) }\theta \left[ \left(
J_1-J\right) ^2-H^2\right] 
\end{equation}
The average value of this expression coincides with Eq.\ref{14}. Thus $%
\left\langle \chi _{eq}\right\rangle $ is generally the unobservable
quantity as it describes the changes of $m$ in the series of transitions
which take place only astronomically large time scale of overbarrier
relaxation. What is more, there are no traces of the transitions in the
limiting expression in Eq.\ref{16} when the number of metastable states goes
to infinity ($N_b\rightarrow \infty $) in the thermodynamic limit. The
indication of their presence via delta functions in $\chi _{eq}$ exists only
when this number stays finite in the limit $N\rightarrow \infty $.

Meanwhile in the framework of the present model $\chi _{eq}$ from Eq.\ref{16}
contains some information on the reaction of inergodic system on the slow
varying external field. Thus the application of slow AC field with amplitude
greater than $2J_1-J$ would give a hysteresis loop and $\chi _{eq}$ defines
its slope. In the same time ordinary measurement of susceptibility in small
fields would give zero value. Possibly the qualitatively similar meaning $%
\chi _{eq}$ has in the real systems. Yet we must note that $\chi _{eq}$ in
present model is a non-self-averaging quantity in the sense that being
constant before averaging it becomes a function of $H$ after it, see Eq.\ref
{14}. Still more dramatic effect is caused by the averaging on the nonlinear
susceptibilities. It follows from Eq.\ref{16} ($k>1$)%
$$
\chi _{eq}^{\left( k\right) }\equiv \frac{\partial ^km_{eq}}{\partial H^k}%
=\frac 1{J_1-J}\frac{\partial ^{k-2}}{\partial H^{k-2}}\left[ \delta \left(
J_1-J+H\right) -\delta \left( J_1-J-H\right) \right] 
$$
and the averaging of this equation gives 
\begin{equation}
\label{17}\left\langle \chi _{eq}^{\left( k\right) }\right\rangle \equiv
\left\langle \frac{\partial ^km_{eq}}{\partial H^k}\right\rangle =-\frac{%
\partial ^{k-2}}{\partial H^{k-2}}\left[ \frac{W\left( \left| H\right|
\right) }H\right] 
\end{equation}
The absence of self-averaging of magnetic susceptibilities seems to be the
specific property of the present model in which small fluctuations of $J_1$
can cause large deviations in $m_{eq}$ and could be absent in more realistic
spin glass models.

Let us note that singularities of non-averaged susceptibilities at $H=\pm
\left( J_1-J\right) $ correspond to the points of the transitions from the
inhomogeneous phase into the homogeneous one, thus $J_1-J$ has the meaning
of the (non-averaged) Almeida-Thouless field. The corresponding anomalies of
the averaged quantities would exist at finite $H=\pm H_{AT}$ if function $%
W\left( J^{\prime }\right) $ in Eq.\ref{12} has a bounded support, i.e. when 
$W\left( J^{\prime }\right) =0$ for $J^{\prime }$ greater some $\overline{J}%
>0$ and $W\left( J^{\prime }\right) >0$ otherwise. Then $H_{AT}=\overline{J}$
and anomalies of $\left\langle \chi _{eq}^{\left( k\right) }\right\rangle $
for $H\rightarrow \pm H_{AT}$ will be determined by the behavior of $W\left(
J^{\prime }\right) $ at $J^{\prime }\rightarrow \overline{J}$. This behavior
determines also how the block Edwards-Anderson order parameter, Eq.\ref{5},
goes to zero when $H\rightarrow \pm H_{AT}$. For $N_b\rightarrow \infty $ we
get from Eq.\ref{5}%
$$
\left\langle q_{eq}\right\rangle =1-\left\langle m_{eq}^2\right\rangle
=\theta \left( \overline{J}-\left| H\right| \right) \int\limits_{\left|
H\right| }^{\overline{J}}dJ^{\prime }W\left( J^{\prime }\right) \left( 1-%
\frac{H^2}{J^{\prime 2}}\right) 
$$

In general case ($W\left( \overline{J}\right) <\infty $) the transition into
spin glass phase with $\left\langle q_{eq}\right\rangle \neq 0$ at $\left|
H\right| <H_{AT}$ is not accompanied with divergencies of $\left\langle \chi
_{eq}^{\left( k\right) }\right\rangle $ in contrast with SK model. This is
because the upturn of the last block along the field is the first order
transition. So the instabilities which could cause such divergencies does
not occur in the present model.

Let us also note that there are two ferromagnetic phases: inergodic phase
with a number of inhomogeneous metastable states and ergodic one having the
unique ferromagnetic state. The average value of the field corresponding the
transition point between these phases can be obtained by considering the
boundaries (upper and lower) of the region where metastable states exist
which represent also the upper and lower branches of hysteresis loop . They
are (see Fig.1(a)):%
$$
m_{\pm }\left( H\right) =\frac{H\pm J_1}{J_1-J}\theta \left[ \left(
J_1-J\right) ^2-\left( H\pm J_1\right) ^2\right] +sign\left( H\pm J_1\right)
\theta \left[ \left( H\pm J_1\right) ^2-\left( J_1-J\right) ^2\right] 
$$
The averaging of this expression over $J_1$ gives 
\begin{equation}
\label{18}\left\langle m_{\pm }\left( H\right) \right\rangle =sign\left(
H\pm J\right) +\theta \left( \mp H-J\right) \int\limits_{\left( \mp
H-J\right) /2}^\infty dJ^{\prime }W\left( J^{\prime }\right) \left( \frac{%
H\pm J}{J^{\prime }}\pm 2\right) 
\end{equation}

For $\left| H\right| $ greater some $H_e$ these branches coincide thus
indicating the transition into ergodic phase. The condition defining $H_e$
is vanishing of the integral in Eq.\ref{18} so $H_e$ will be finite when $%
W\left( J^{\prime }\right) $ has a bounded support. In this case%
$$
H_e=2\overline{J}+J=2H_{AT}+J 
$$
There exists a functional relation between $\left\langle m_{\pm }\left(
H\right) \right\rangle $ and $\left\langle m_{eq}\left( H\right)
\right\rangle $, Eq.\ref{13} of the following form: 
\begin{equation}
\label{19}\left\langle m_{\pm }\left( H\right) \right\rangle =\mp \left[
2\theta \left( \mp H-J\right) \left\langle m_{eq}\left( \frac{\mp H-J}%
2\right) \right\rangle -1\right] 
\end{equation}
From Eq.\ref{19} follows also: 
\begin{equation}
\label{20}\frac{\partial \left\langle m_{\pm }\left( H\right) \right\rangle 
}{\partial H}=\theta \left( \mp H-J\right) \left\langle \chi _{eq}\left( 
\frac{\mp H-J}2\right) \right\rangle 
\end{equation}
These relations are specific for the model under consideration but,
probably, some relations between field dependency of the average equilibrium
magnetization and hysteresis loop contour exist also in other spin glass
models at $T=0$.

For the simple 'rectangular' function $W$ 
$$
W\left( J^{\prime }\right) =\theta \left( J^{\prime }-\overline{J}\right) /%
\overline{J} 
$$
we get%
$$
\left\langle q_{eq}\right\rangle =\theta \left( \overline{J}^2-H^2\right)
\left( 1-\frac{\left| H\right| }{\overline{J}}\right) ^2 
$$
$$
\left\langle m_{eq}\left( H\right) \right\rangle =\theta \left( \overline{J}%
^2-H^2\right) \frac H{\overline{J}}\ln \left( \frac{e\overline{J}}{\left|
H\right| }\right) +\theta \left( H^2-\overline{J}^2\right) 
$$
$$
\overline{J}\left\langle \chi _{eq}\right\rangle =\theta \left( \overline{J}%
^2-H^2\right) \ln \left( \frac{\overline{J}}{\left| H\right| }\right) 
$$
$$
\left\langle \chi _{eq}^{\left( k\right) }\right\rangle =\left( k-1\right)
!\left( -H\right) ^{1-k}/\overline{J} 
$$
The divergency of magnetic susceptibilities at $H=0$ is a consequence of $%
W\left( 0\right) \neq 0$. They would be finite at zero field if $W$ goes to
zero as some power of $J^{\prime }$ or faster when $J^{\prime }\rightarrow 0$%
. For example, the averaging with the 'triangle' function 
\begin{equation}
\label{21}W\left( J^{\prime }\right) =2J^{\prime }\theta \left( J^{\prime }-%
\overline{J}\right) /\overline{J^2} 
\end{equation}
gives the following results%
$$
\left\langle q_{eq}\right\rangle =\theta \left( \overline{J}^2-H^2\right)
\left( 1+\frac{H^2}{\overline{J}^2}\ln \frac{H^2}{e\overline{J}^2}\right) 
$$
$$
\left\langle m_{eq}\left( H\right) \right\rangle =\theta \left( \overline{J}%
^2-H^2\right) \frac H{\overline{J}}\left( 2-\frac{\left| H\right| }{%
\overline{J}}\right) +\theta \left( H^2-\overline{J}^2\right) signH 
$$
$$
\overline{J}\left\langle \chi _{eq}\right\rangle =\theta \left( \overline{J}%
^2-H^2\right) 2\left( 1-\frac{\left| H\right| }{\overline{J}}\right) 
$$
$$
\left\langle \chi _{eq}^{\left( k\right) }\right\rangle =0,k>1 
$$
The averaging of the thermodynamic parameters of metastable states existing
inside hysteresis loop is trivial their magnetizations do not depend on $H$
and susceptibilities are zero. Also $q=1-m^2$ and $S=0$ in all states. But
we must note that equilibrium entropy is not strictly zero as for a given $m$
there are $\binom{N_b}{N_b\left( 1-m\right) /2}$ states with equal
potentials so the configurational entropy term%
$$
S_{conf}\left( m\right) =N^{-1}\ln \binom{N_b}{N_b\left( 1-m\right) /2} 
$$
is added to the expression in Eq.\ref{6}. But in the thermodynamic limit $%
S_{conf}\left( m\right) $ goes to zero as $N_s^{-1}$.

\section{Thermodynamics near the transition.}

The stable inhomogeneous solutions of equations of state, Eq.\ref{7}, appear
at $T<J_1$. So in the case considered here ($J_1>J$) there is a second order
phase transition from homogeneous paramagnetic phase into inhomogeneous
inergodic spin glass one at $T=J_1,H=0$. Let us consider the thermodynamics
of the model in the vicinity of this transitions assuming 
\begin{equation}
\label{22}m_\alpha <<1 
\end{equation}
In this case Eq.\ref{7} acquire the form: 
\begin{equation}
\label{23}\tau _1m_\alpha +\left( \tau -\tau _1\right) m+m_\alpha ^3/3=h 
\end{equation}
Here $h=H/T,\tau _1=1-J_1/T,\tau =1-J/T,\tau _1>\tau $. Hessian, Eq.\ref{8},
becomes in this region 
\begin{equation}
\label{24}T^{-1}\frac{\partial ^2F\left( {\bf m}\right) }{\partial m_\alpha
\partial m_\beta }=\left( \tau _1+m_\alpha ^2\right) \delta _{\alpha \beta
}+\left( \tau -\tau _1\right) /N 
\end{equation}
It follows from Eq.\ref{22} and Eq.\ref{23} that%
$$
h<<1,\tau _1<<1,\tau <<1 
$$
For these conditions to be fulfilled for every random $J_1$ we must assume
that $W\left( J^{\prime }\right) $ in Eq.\ref{12} has sufficiently narrow
bounded support, that is the possible values of $J_1-J$ must be less than
some $\overline{J}>0$ obeying the condition%
$$
\overline{J}<<J 
$$
When $\tau _1>0$ Eqs.\ref{23} have unique homogeneous solution. Let us
denote it as $m_0$. It does not depend on $\tau _1$ and obeys the equation 
\begin{equation}
\label{25}\tau m_0+m_0^3/3=h 
\end{equation}
When $\tau _1<0$ Eqs.\ref{23} have $3^{N_b}-3$ inhomogeneous solutions
beside $m_0$. As all $m_\alpha $ obey the same cubic equation they can
acquire only three different values which we denote as $\widetilde{m}%
_s,s=1,2,3$. Then all inhomogeneous solutions can be characterized by the
three numbers $n_s\neq N_b$,%
$$
\sum_sn_s=N_b 
$$
which show how many $m_\alpha $ have the value $\widetilde{m}_s$. There are $%
\frac{N_b}{n_1!n_2!n_3!}$ solutions which differ by the permutations of $%
m_\alpha $ and the total number of solutions is%
$$
\sum_{n_s\neq N_b}\frac{N_b}{n_1!n_2!n_3!}=3^{N_b}-3 
$$
But only $2^{N_b}-2$ of them could be stable. Indeed, Hessian, Eq.\ref{24},
has three eigenvalues equal to $\tau _1+\widetilde{m}_s^2$ with degeneracy $%
n_s-1$, which correspond to the eigenvectors having a zero sum of
components. There are also three non-degenerate eigenvalues which are the
solutions of the equation 
\begin{equation}
\label{26}1+\frac{\tau -\tau _1}{N_b}\sum_s\frac{n_s}{\tau _1+\widetilde{m}%
_s^2-\lambda }=0 
\end{equation}
Using the Viet's theorem for Eq.\ref{23} according to which 
\begin{equation}
\label{27}\sum_s\widetilde{m}_s=0 
\end{equation}
\begin{equation}
\label{28}\sum_{s<s^{\prime }}\widetilde{m}_s\widetilde{m}_{s^{\prime
}}=3\tau _1 
\end{equation}
we can get the relation%
$$
\sum_s\frac 1{\tau _1+\widetilde{m}_s^2}=0 
$$
It shows that all three eigenvalues $\tau _1+\widetilde{m}_s^2$ could not be
positive simultaneously so the stable solutions must have at least one of
the numbers $n_s$ equal to 0 or 1. But if all $n_s>0$ then one of
the solutions of Eq.\ref{26} becomes negative for large $N_b>>1$. Thus the
stable solutions must have one of $n_s$ equal to zero. Further we will
consider just these solutions putting $n_3=0$. The stability condition for
them reduces to one inequality 
\begin{equation}
\label{29}\tau _1+\widetilde{m}_3^2<0 
\end{equation}
It follows from Eq.\ref{27} and Eq.\ref{28} that $\widetilde{m}_s$ can be
represented in the following form%
$$
\widetilde{m}_1=2\left( -\tau _1\right) ^{1/2}\cos \left( \varphi -\frac \pi
6\right) 
$$
$$
\widetilde{m}_2=-2\left( -\tau _1\right) ^{1/2}\cos \left( \varphi +\frac
\pi 6\right) 
$$
\begin{equation}
\label{30}\widetilde{m}_3=-2\left( -\tau _1\right) ^{1/2}\sin \varphi 
\end{equation}
so the stability condition, Eq.\ref{29}, is equivalent to the inequality 
\begin{equation}
\label{31}\left| \varphi \right| <\pi /6 
\end{equation}
It follows from the definition of $m$:%
$$
m=\nu _1\widetilde{m}_1+\nu _2\widetilde{m}_2 
$$
Here 
\begin{equation}
\label{32}\nu _s=\frac{n_s}{N_b} 
\end{equation}
$$
\nu _1+\nu _2=1 
$$
so 
\begin{equation}
\label{33}m=\left( -\tau _1\right) ^{1/2}\left( \sqrt{3}\Delta \cos \varphi
+\sin \varphi \right) 
\end{equation}
$$
\Delta =\nu _1-\nu _2 
$$
Inserting Eq.\ref{30} into Eq.\ref{23} we get 
\begin{equation}
\label{34}3\left( \tau -\tau _1\right) m=3h-2\left( -\tau _1\right)
^{3/2}\sin 3\varphi 
\end{equation}
Excluding $m$ from Eq.\ref{33} and Eq.\ref{34} we obtain the equation for $%
\varphi $: 
\begin{equation}
\label{35}2\left( -\tau _1\right) ^{3/2}\sin 3\varphi +3\left( \tau -\tau
_1\right) \left( -\tau _1\right) ^{1/2}\left( \sqrt{3}\Delta \cos \varphi
+\sin \varphi \right) =3h 
\end{equation}
At all $\tau >\tau _1$ the left side of Eq.\ref{35} is a monotonously
growing function of $\varphi $ for $\left| \varphi \right| <\pi /6$. Hence,
there is only one stable solution for $\widetilde{m}_s$ at a given $\Delta $%
. There are $\binom{N_b}{n_1}$ metastable states corresponding to this
solution which differ by $m_\alpha $ permutations.

The explicit solution of Eq.\ref{35} can be found for $\Delta =0$ when it
becomes cubic. In the limiting cases $\Delta =\pm 1$ Eq.\ref{35} reduces
also to a cubic one for $\widetilde{m}_1$ or $\widetilde{m}_2$ which
coincides with Eq.\ref{25}.

In general case Eq.\ref{35} and Eq.\ref{33} (or Eq.\ref{34}) give a
parametric representation of a dependency of the homogeneous magnetization
in the metastable states with a given $\Delta $ on $\tau $, $\tau _1$ and $h$%
. The parameter $\varphi $ can be excluded from these equations with the
result 
$$
\left[ \left( 3\Delta ^2+1\right) ^2\tau -3\left( 1-3\Delta ^2\right) \left(
1-\Delta ^2\right) \tau _1\right] m+\frac{8\left( 9\Delta ^2-1\right) }{%
3\left( 3\Delta ^2+1\right) }m^3+ 
$$
\begin{equation}
\label{36}+2\sqrt{3}\frac{\Delta \left( 1-\Delta ^2\right) }{3\Delta ^2+1}%
\left[ \left( 3\Delta ^2+1\right) \tau _1+4m^2\right] \left[ -\left( 3\Delta
^2+1\right) \tau _1-m^2\right] ^{1/2}=\left( 3\Delta ^2+1\right) ^2h 
\end{equation}
From the stability condition, Eq.\ref{31}, and Eq.\ref{34} it follows that
solutions of Eq.\ref{36} is stable in the region 
\begin{equation}
\label{37}9\left[ \left( \tau -\tau _1\right) m-h\right] ^2<-4\tau _1^3 
\end{equation}
which is the band on the $m-h$ plane. The magnetization is a monotonously
growing function of $h$ and $\Delta $ inside this band so the field
dependencies of magnetization can be represented as a set of uncrossing
lines bounded from above and below by the $m_0(h)$ line as shown in Fig.2.%
\begin{figure}
\centering
\includegraphics[160,335][435,570]{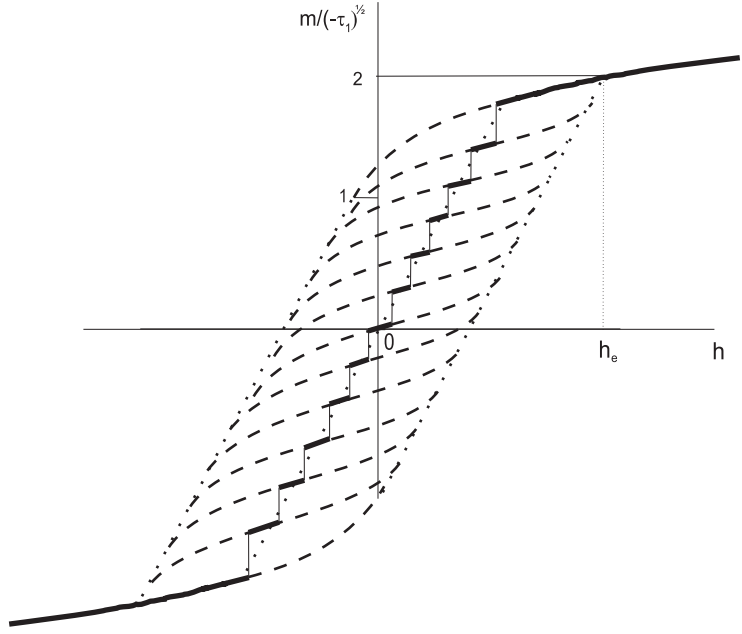}
\caption{Field dependencies of magnetization in
metastable states (dashed lines) and stable states (solid lines) near
transition, $\tau<0$.}
\end{figure}
The other thermodynamic parameters of metastable states can be obtained by
differentiation of thermodynamic potential which near transition has the
form 
\begin{equation}
\label{38}12F/T=6\left( \tau -\tau _1\right) m^2+6\tau _1\sum_s\nu _s%
\widetilde{m}_s^2+\sum_s\nu _s\widetilde{m}_s^4-12hm-12\ln 2 
\end{equation}
Expressed via $\varphi $ these parameters are%
$$
q=3\tau _1\left( \Delta ^2-1\right) \cos \varphi 
$$
$$
T^{-1}\chi ^{-1}=\tau -\tau _1\left[ 1+2\cos 3\varphi /\left( \cos \varphi
-\sqrt{3}\Delta \sin \varphi \right) \right] 
$$
$$
S=\ln 2+\tau _1\left( 2+\cos 2\varphi +\sqrt{3}\Delta \sin 2\varphi \right)
/2 
$$
For the heat capacity we get rather more cumbersome expression 
$$
C=\frac{2\tau _1\left(\sqrt{3}\Delta \sin \varphi -\cos \varphi -2\cos
3\varphi \right) +3\left( \tau -\tau _1\right) \left( 1-\Delta ^2\right)
\cos \varphi }{2\left( \cos \varphi -\sqrt{3}\Delta \sin \varphi \right)
-4\tau _1\cos \varphi } 
$$
In spite of the absence of explicit expression for $\varphi $ as a function
of $h$, $\tau $ and $\tau _1$ the above formulae allow to get some notion
about the field and temperature dependencies of these quantities. Thus at
the boundaries of stability region, $\left| \varphi \right| =\pm \pi /6$ or
at 
\begin{equation}
\label{39}h=\left( -\tau _1\right) ^{1/2}\left( \tau -\tau _1\right) \left(
3\Delta \pm 1\right) /2\pm 2\left( -\tau _1\right) ^{3/2}/3 
\end{equation}
$q$ and $\chi ^{-1}$ has the lowest values%
$$
\chi ^{-1}=J_1-J 
$$
\begin{equation}
\label{40}q=9\tau _1\left( \Delta ^2-1\right) /4 
\end{equation}
while the entropy and heat capacity are%
$$
S=\ln 2+3\tau _1\left( 1\pm \Delta \right) /2 
$$
\begin{equation}
\label{41}C=3\left( 1\pm \Delta \right) /2-\tau _1/\left( \tau -\tau
_1\right) 
\end{equation}
It follows from Eq.\ref{39} that metastable states exist when 
\begin{equation}
\label{42}\left| h\right| <\widetilde{h}_e\equiv 2\left( -\tau _1\right)
^{1/2}\left( \tau -\frac 43\tau _1\right) 
\end{equation}
When $h$ goes to $\pm h_e$ the more homogeneous states with $\Delta
\rightarrow \pm 1$ stay stable and their magnetization tends to $m_0\left( 
\widetilde{\pm h}_e\right) =\pm 2\left( -\tau _1\right) ^{1/2}$. However the
limiting values of magnetic susceptibility, entropy and heat capacity differ
from those in homogeneous state: $\chi _0^{-1}=J_1\left( \tau +m_0^2\right) $%
, $S_0=\ln 2-m_0^2/2$, $C=m_0^2/\left( \tau +m_0^2\right) $.

In the middle of the stability band (at $\varphi =0$ or $h=\Delta \left(
-3\tau _1\right) ^{1/2}\left( \tau -\tau _1\right) $) we get: $q=\tau
_1\left( \Delta ^2-1\right) $, $\chi ^{-1}=J_1\left( 1-2\tau _1\right) -J$, $%
S=\ln 2+3\tau _1/2$, $C=\frac 32\left( 1-\Delta ^2\frac{\tau -\tau _1}{\tau
-3\tau _1}\right) $. In this case the diminishing of inhomogeneity when $%
h\rightarrow \pm \widetilde{h}_{AT}$, 
\begin{equation}
\label{43}\widetilde{h}_{AT}=\left( -3\tau _1\right) ^{1/2}\left( \tau -\tau
_1\right) 
\end{equation}
$m$, $\chi $, $S$ and $C$ tend to the corresponding values of the
homogeneous phase.

The Almeida-Thouless field $\widetilde{h}_{AT}$, Eq. \ref{43} determines (to
the order $N_b^{-1}$) the point of the transition into the homogeneous
phase. To show this let us find the values $\Delta _{eq}$ corresponding to
the states with the lowest potential. Differentiating $F$ , Eq.\ref{38} over 
$\Delta $ and using Eqs.\ref{23}, \ref{27} and Eq.\ref{28} we get%
$$
\frac{\partial F}{\partial \Delta }=T\widetilde{m}_3\left( \widetilde{m}_1-%
\widetilde{m}_2\right) ^3/24 
$$
$$
\frac{\partial ^2F}{\partial \Delta ^2}=\frac{T\left( \tau -\tau _1\right)
\left( \widetilde{m}_1-\widetilde{m}_2\right) ^2}{8\left[ 1+\sum\limits_s%
\frac{\nu _s}{\tau +\widetilde{m}_s^2}\right] } 
$$
Thus the states with $\widetilde{m}_3=0$ or, equivalently, $\varphi =0$ (cf.
Eq.\ref{30}) have the lowest potential. One can see that Eq.\ref{35} has
solution $\varphi =0$ when $\Delta =h/\widetilde{h}_{AT}$ which is possible
at $h^2<\widetilde{h}_{AT}^2$. When $h^2>\widetilde{h}_{AT}^2$ $F\left(
\Delta \right) $ has no minima inside the region $\Delta ^2<1$ in which it
is defined and the minimal values occur at its boundaries for $\Delta
_{eq}=signH$. So the transition into homogeneous state takes place at $h=\pm 
\widetilde{h}_{AT}$.

As $\Delta $ is a rational number of the form $2n/N_b-1$ (cf. Eq.\ref{32})
it can not be exactly equal to $h/\widetilde{h}_{AT}$ at all $h^2<\widetilde{%
h}_{AT}^2$. Hence $\Delta _{eq}$ is defined so that $\left| \Delta -h/%
\widetilde{h}_{AT}\right| $ is minimal and can be represented as%
$$
\Delta _{eq}=\sum\limits_{n=1}^{N_b-1}\left( \frac{2n}{N_b}-1\right) \theta
\left( N_b^{-2}-\widetilde{\varepsilon }_n^2\right) +signH\theta \left[
h^2-\left( \frac{N_b-1}{N_b}\right) \widetilde{h}_{AT}^2\right] 
$$
$$
\widetilde{\varepsilon }_n\equiv \frac h{\widetilde{h}_{AT}}-\frac{2n}{N_b}%
+1 
$$
Inserting this $\Delta _{eq}$ into Eq.\ref{35} we get the corresponding
values of $\varphi _{eq}$ at $h^2<\widetilde{h}_{AT}^2$:%
$$
\varphi _{eq}=\sqrt{3}\frac{\tau -\tau _1}{\tau -3\tau _1}\sum%
\limits_{n=1}^{N_b-1}\widetilde{\varepsilon }_n\theta \left( N_b^{-2}-%
\widetilde{\varepsilon }_n^2\right) 
$$
Inserting $\Delta _{eq}$ and $\varphi _{eq}$ into the parametric
representations of $q$ and $m$ we obtain the equilibrium values of these
quantities%
$$
q_{eq}=-3\tau _1\left( 1-h^2/\widetilde{h}_{AT}^2\right) 
$$
$$
m_{eq}=\frac h{\tau -\tau _1}\theta \left[ \left( \frac{N_b-1}{N_b}\right) 
\widetilde{h}_{AT}^2-h^2\right] -\frac{2\sqrt{3}\left( -\tau _1\right) ^{3/2}%
}{\tau -3\tau _1}\sum\limits_{n=1}^{N_b-1}\widetilde{\varepsilon }_n\theta
\left( N_b^{-2}-\widetilde{\varepsilon }_n^2\right) + 
$$
$$
+m_0\theta \left[ h^2-\left( \frac{N_b-1}{N_b}\right) \widetilde{h}%
_{AT}^2\right] 
$$
Differentiating $m_{eq}$ over $h$ we get the equilibrium susceptibility%
$$
\chi _{eq}=\frac h{\tau -3\tau _1}\theta \left[ \left( \frac{N_b-1}{N_b}%
\right) \widetilde{h}_{AT}^2-h^2\right] -\frac{4\tau _1}{N_b\left( \tau
-\tau _1\right) \left( \tau -3\tau _1\right) }\sum\limits_{n=1}^{N_b-1}%
\delta \left( N_b^{-1}-\widetilde{\varepsilon }_n\right) + 
$$
$$
+\frac 1{\tau +m_0^2}\theta \left[ h^2-\left( \frac{N_b-1}{N_b}\right) 
\widetilde{h}_{AT}^2\right] 
$$
The equilibrium entropy can be obtained by the differentiation of the
equilibrium potential which to the $\widetilde{\varepsilon }_n^2$ order is%
$$
F_{eq}=F\left( \Delta =h/\widetilde{h}_{AT}\right) -TS_{conf} 
$$
where configurational entropy $S_{conf}$ is determined by the logarithm of
the number of states with the same potential $F$,%
$$
S_{conf}=N^{-1}\ln \binom{N_b}{N_b\left( 1-\Delta _{eq}\right) /2} 
$$
As at $T=0$, $S_{conf}$ is of the order $N_s^{-1}$and can be neglected so%
$$
S_{eq}=\ln 2+\frac 32\tau _1\theta \left[ \left( \frac{N_b-1}{N_b}\right) 
\widetilde{h}_{AT}^2-h^2\right] -\frac{2\sqrt{3}\left( -\tau _1\right) ^{1/2}h%
}{\tau -3\tau _1}\sum\limits_{n=1}^{N_b-1}\widetilde{\varepsilon }_n\theta
\left( N_b^{-2}-\widetilde{\varepsilon }_n^2\right) - 
$$
$$
-\frac{m_0^2}2\theta \left[ h^2-\left( \frac{N_b-1}{N_b}\right) \widetilde{h}%
_{AT}^2\right] 
$$
For the equilibrium heat capacity we get%
$$
C_{eq}=\left( \frac 32+\frac{h^2}{\tau _1\left( \tau -\tau _1\right) \left(
\tau -3\tau _1\right) }\right) \theta \left[ \left( \frac{N_b-1}{N_b}\right) 
\widetilde{h}_{AT}^2-h^2\right] - 
$$
$$
-\frac{2h^2}{N_b\tau _1\left( \tau -\tau _1\right) \left( \tau -3\tau
_1\right) }\sum\limits_{n=1}^{N_b-1}\delta \left( N_b^{-1}-\widetilde{%
\varepsilon }_n\right) +\frac{m_0^2}{\tau +m_0^2}\theta \left[ h^2-\left( 
\frac{N_b-1}{N_b}\right) \widetilde{h}_{AT}^2\right] 
$$
The averaging of these expressions over $J_1$ gives at large $N_b$%
$$
\left\langle q_{eq}\right\rangle =\int\limits_{H/m_0}^{\overline{J}%
}dJ^{\prime }W\left( J^{\prime }\right) \left( 3\frac{J^{\prime }}J-3\tau -%
\frac{H^2}{J^{\prime 2}}\right) 
$$
$$
\left\langle m_{eq}\right\rangle =\int\limits_{H/m_0}^{\overline{J}}\frac{%
dJ^{\prime }}{J^{\prime }}W\left( J^{\prime }\right)
+m_0\int\limits_0^{H/m_0}dJ^{\prime }W\left( J^{\prime }\right) 
$$
$$
\left\langle \chi _{eq}\right\rangle =\int\limits_{H/m_0}^{\overline{J}}%
\frac{dJ^{\prime }}{J^{\prime }}W\left( J^{\prime }\right) +\frac 1{J\left(
\tau +m_0^2\right) }\int\limits_0^{H/m_0}dJ^{\prime }W\left( J^{\prime
}\right) 
$$
$$
\left\langle S_{eq}\right\rangle =\ln 2+\frac 32\int\limits_{H/m_0}^{%
\overline{J}}dJ^{\prime }W\left( J^{\prime }\right) \left( \tau -\frac{%
J^{\prime }}J\right) -\frac{m_0^2}2\int\limits_0^{H/m_0}dJ^{\prime }W\left(
J^{\prime }\right) 
$$
$$
\left\langle C_{eq}\right\rangle =\frac 32\int\limits_{H/m_0}^{\overline{J}%
}dJ^{\prime }W\left( J^{\prime }\right) +\frac{m_0^2}{\tau +m_0^2}%
\int\limits_0^{H/m_0}dJ^{\prime }W\left( J^{\prime }\right) 
$$
In derivation of these expression we have used the equivalence of the
condition $h^2<\widetilde{h}_{AT}^2$ and the inequality $m_0^2\left(
h\right) <m_0^2\left( h_{AT}\right) =-3\tau _1$ where $m_0\left( h\right) $
is a solution of Eq.\ref{25} such as $m_0\left( h\right) h>0$. In its turn,
from Eq.\ref{25} it follows that the last inequality is equivalent to $%
J_1-J>H/m_0$.

Evidently, the average equilibrium parameters transfer continuously into
corresponding values of homogeneous phase at $H>\overline{J}m_0$. Thus the
average Almeida-Thouless field $H_{AT}$ is defined as a solution of the
equation $H_{AT}=\overline{J}m_0\left( H_{AT}\right) $ or its equivalent 
\begin{equation}
\label{44}m_0^2\left( H_{AT}\right) =\sigma 
\end{equation}
where $\sigma =\overline{J}/T-\tau =\left( J+\overline{J}\right) /T-1$.

Let us remind that from Eq.\ref{22} it follows that $J>>\overline{J}$. The
solution of Eq.\ref{44} exists when $\sigma >0$ or $T>T_{sg}=J+\overline{J}$%
. For $T\rightarrow T_{sg}$, $\sigma <<\overline{J}/J$ we get%
$$
H_{AT}\approx \overline{J}\left( 3\sigma \right) ^{1/2} 
$$
and for $\tau <0$, $-\tau >>\overline{J}/J$%
$$
H_{AT}\approx \overline{J}\left( -3\tau \right) ^{1/2} 
$$
Let us further consider the average boundaries of inergodic region on $m-h$
plane, i.e. the branches of the average hysteresis loop 
$$
\left\langle m_{\pm }\left( H\right) \right\rangle =m_0\theta \left(
m_0^2-4\sigma \right) +m_0^{\pm }\theta \left( -\tau \right) \theta \left(
\pm h_{\pm }\right) \theta \left[ 4\sigma -\left( m_0^{\pm }\right)
^2\right] \theta \left[ \left( m_0^{\pm }\right) ^2-\sigma \right] + 
$$
\begin{equation}
\label{45}+\theta \left( \pm h_{\pm }\right) \theta \left[ \sigma -\left(
m_0^{\pm }\right) ^2\right] A_{\pm }\left[ \tau +\left( m_0^{\pm }\right)
^2\right] +\theta \left( \mp h_{\pm }\right) \theta \left[ 4\sigma -\left(
m_0^{\pm }\right) ^2\right] A_{\pm }\left( \tau +m_0^2/4\right) 
\end{equation}
Here $h_{\pm }=h\pm \theta \left( -\tau \right) 2\left( -\tau \right)
^{3/2}/3$, 
$$
A_{\pm }\left( z\right) =J\int\limits_z^{\overline{J}/T}\frac{dx}xW\left(
Jx\right) \left[ h\pm \frac 23\left( x-\tau \right) ^{3/2}\right] +Jm_0^{\pm
}\int\limits_0^zdxW\left( Jx\right) 
$$
and $m_0^{+}$ and $m_0^{-}$ are the maximal and minimal solutions of Eq.\ref
{25} correspondingly which exist at $\tau <0$, $4\tau +m_0^2<0$. When $\tau
>0$ or $\tau <0$, $4\tau +m_0^2>0$ then $m_0^{\pm }=m_0$.

Eq.\ref{45} shows that the branches of the average hysteresis loop coincide
when $H^2>H_e^2$, $H_e$ being the solution of equation%
$$
m_0^2\left( H_e\right) =4\sigma 
$$
For $T\rightarrow T_{sg}$, $\sigma <<\overline{J}/J$ we get%
$$
H_e\approx 2\overline{J}\sigma ^{1/2} 
$$
and for $\tau <0$, $-\tau >>\overline{J}/J$, $H_e$ almost coincide with the
coercive field for the homogeneous solution%
$$
H_e\approx 2J\left( -\tau \right) ^{3/2}/3 
$$
so the hysteresis loops becomes similar to that of ordinary ferromagnet.

Let us present the explicit expressions for the functions $A_{\pm }$ in Eq.%
\ref{45} for the 'triangular' function $W$, Eq.\ref{21},:%
$$
A_{\pm }\left[ \tau +\left( m_0^{\pm }\right) ^2\right] =\frac{J^2}{%
\overline{J}^2}\left[ 2h\sigma \pm \frac 8{15}\sigma ^{5/2}+m_0^{\pm }\left(
\tau ^2-\frac 15\left( m_0^{\pm }\right) ^4\right) \right] 
$$
$$
A_{\pm }\left[ \tau +m_0^2/4\right] =\frac{J^2}{\overline{J}^2}\left[
2h\sigma \pm \frac 8{15}\sigma ^{5/2}+m_0\left( \tau ^2-\frac
7{80}m_0^4\right) \right] 
$$
For the same $W$ the average equilibrium parameters at $H^2<H_{AT}^2$ are%
$$
\left\langle q_{eq}\right\rangle =2\sigma \left[ 1-\left( \frac H{m_0%
\overline{J}}\right) ^3\right] -\tau \left[ 1-3\left( \frac H{m_0\overline{J}%
}\right) ^2+2\left( \frac H{m_0\overline{J}}\right) ^3\right] +2\left( \frac
H{\overline{J}}\right) ^2\ln \left( \frac H{m_0\overline{J}}\right) 
$$
$$
\left\langle m_{eq}\right\rangle =2\frac H{\overline{J}}-\frac{H^2}{m_0%
\overline{J}^2} 
$$
$$
\left\langle \chi _{eq}\right\rangle =\frac 2{\overline{J}}-\frac{2H}{m_0%
\overline{J}^2}+\frac{H^2}{m_0^2\overline{J}^2J\left( \tau +m_0^2\right) } 
$$
$$
\left\langle S_{eq}\right\rangle =\ln 2-\sigma \left[ 1-\left( \frac H{m_0%
\overline{J}}\right) ^3\right] +\frac \tau 2\left[ 1-3\left( \frac H{m_0%
\overline{J}}\right) ^2+2\left( \frac H{m_0\overline{J}}\right) ^3\right]
-\frac 12\left( \frac H{\overline{J}}\right) ^2 
$$
$$
\left\langle C_{eq}\right\rangle =\frac 32\left[ 1-\left( \frac H{m_0%
\overline{J}}\right) ^2\right] +\frac{H^2}{\overline{J}^2\left( \tau
+m_0^2\right) } 
$$

Let us note once more that the average equilibrium parameters are generally
unobservable quantities. Probably, the experimental values being rather
close to them are obtained after cooling in small external fields
(field-cooled (FC) regime) for $T$ near $T_{sg}$ \cite[17]{13} when barriers
between metastable states are relatively small and system could relax into
the lowest (or close to it) state at a sufficiently slow cooling. In zero
field cooled (ZFC) regime when field is applied after cooling below $T_{sg}$
in zero field, the observed thermodynamical parameters would differ from
equilibrium ones as the system would at first be trapped in the state with $%
\Delta =0$ and will stay in it if applied field does not exceed $h_c=\left(
-\tau _1\right) ^{1/2}\left( \tau -4\tau _1/3\right) /2$, cf. Eq.\ref{39}.
Thus at $h<h_c$ the ZFC parameters are those of $\Delta =0$ metastable
states. When applied field $h>h_c$ the system relaxes into the metastable
state at the boundary of stability region (on the lower branch of hysteresis
loop) having some $\Delta >0$ which is a solution of Eq.\ref{39}. Inserting
this $\Delta $ in Eq.\ref{40} and Eq.\ref{41} we get the values of
thermodynamic parameters the observed quantities would relax to in ZFC
regime at $h>h_c$.

Similarly, the parameters of metastable states define the other quantities
which are determined in the slow nonequilibrium processes in the spin glass
phase, such as thermo-remanent magnetization, $m_{TRM}$, which remains after
FC process and subsequent switching off the field, and isothermal remanent
magnetization, $m_{IRM}$, remaining after ZFC process followed by the
application for some time (longer than the intravalley relaxation time) an
external field \cite[17]{13}. Thus $m_{IRM}$ is apparently nonzero only at $%
h>h_c$ and an equation defining it can be obtained by putting $h=0$ in Eq.%
\ref{36} and inserting in this equation the value of $\Delta $ we get from
Eq.\ref{39}. The equation for $m_{IRM}$ can also be obtained from Eq.\ref{36}
by putting in it $h=0$ and $\Delta =\min \left( 1,h/h_{AT}\right) $.

\section{Conclusions}

The most remarkable feature of the model considered is the possibility to
imitate the properties of such complex systems as spin glass with the aid of
very simple Hamiltonian. It is common belief that the existence of a number
of metastable states in spin glasses is caused by the frustration of random
Hamiltonian, that is the absence of unique spin configuration providing the
minimal energy \cite{13}. In the present model the degeneracy of the ground
state results from the permutational symmetry of the Hamiltonian instead of
frustration. Nevertheless there also exists the transition into the
inergodic phase and its magnetic properties appear to be very similar to
those of real spin glasses including a set of transitions between metastable
states and inclined hysteresis loop resulting from their presence \cite[17]
{13}. One may suppose that more realistic random Hamiltonians can also have
some approximate permutational symmetry and corresponding quasidegeneracy of
ground state more essential, perhaps, than that caused by the frustration.

We may note that present model implies the definite mechanism of the
transitions between metastable states in a field, namely, the spin-flop
transitions caused by the antiferromagnetic interaction between
macroscopically large spin blocks. It seems rather probable that in some
Ising short-range models of spin glasses it is possible to distinguish many
clusters with the mostly antiferromagnetic interactions at the boundaries
and relatively weak interactions inside them. Still it is rather evident
that there can be also a many more other bond configurations in which
degenerate spin configurations can exist at some field values. Thus it is
hard to say to what extent spin-flop transitions are characteristic for real
spin glasses and if the Hamiltonian, Eq.\ref{4}, is a reasonable
approximation for some random short-range Hamiltonian with some specific
type of disorder. Still such possibility seems to be rather probable in view
of similarity of the properties of the model to those of some real
disordered magnets \cite[17]{13}.

Finally we may state that in spite of the qualitative nature of the model it
allows to get some notion about the character of theoretical results
relevant for the description of real experiments in inergodic systems. It
shows how these result may look like and how the thermodynamic parameters of
metastable states are related to the characteristics of nonequilibrium
processes in spin glass phases.

This work was made under support from Russian Foundation for Basic
Researches, Grants N 98-02-18069 and 97-02-17878.

\end{document}